\documentclass{ws-procs975x65}

\begin{document}

\title{``Singularity" of Levi-Civita Spacetimes}

\author{D. A. KONKOWSKI}

\address{Department of Mathematics, \\
U.S. Naval Academy, \\ 
Annapolis, Maryland, 21402, USA\\ 
E-mail: dak@usna.edu}

\author{T. M. HELLIWELL and C. WIELAND }

\address{Department of Physics, \\ 
Harvey Mudd College, \\
Claremont, California, 91711, USA\\
E-mail: T\_Helliwell@HMC.edu}

\maketitle

\abstracts{
Levi-Civita spacetimes have both classical and quantum singularities. The relationship between the two is
used here to study and clarify the physical aspects of the enigmatic Levi-Civita 
spacetimes.}

\section{Introduction}
Classically singular spacetimes may be quantum mechanically nonsingular. Whereas 
classical singularities are indicated by incomplete geodesics or incomplete paths of
bounded acceleration in maximal spacetimes [\refcite{HE}, \refcite{ES}], quantum singularities are indicated
by quantum wave packets whose behavior is not completely defined by the wave equation
and the underlying spacetime. In other words, spacetimes are quantum mechanically
singular if boundary conditions need to be introduced at the classical singularity to uniquely
specify the quantum wave behavior. Technically, quantum mechanically singular spacetimes are those in which the spatial derivative
operator in a wave equation such as the Klein-Gordon equation is not essentially
self adjoint on a $C_{0}^{\infty}$ domain in $L^{2}$, a Hilbert space of square integrable functions. Quantum singularities were first considered by 
Horowitz and Marolf [\refcite{HM}] following earlier work by Wald [\refcite{Wald}]. Certain
classically singular spacetimes are quantum mechanically nonsingular
(e.g., certain orbifolds and extreme Kaluza-Klein [\refcite{HM}]), but other classically
singular spacetimes are still singular when probed by quantum wave packets
(e.g., Reissner-Nordstr\"{o}m, negative mass Schwarzschild and various quasiregular spacetimes [\refcite{HM}, \refcite{HK}]). Here we use a physically insightful method known as
Weyl's limit point-limit circle criterion [\refcite{RS}] to study
the quantum singularities in Levi-Civita spacetimes and gain insight into the meaning
of the metric parameters of these enigmatic spacetimes. 
This conference proceeding is based on [\refcite{KHW}].

\section{Levi-Civita Spacetimes}

The metric for a Levi-Civita spacetime [\refcite{LC}] has the form

\begin{equation}
    ds^{2} = r^{4 \sigma}dt^{2} - r^{8 \sigma^{2} - 4\sigma}(dr^{2}+
    dz^{2}) - \frac{r^{2 - 4 \sigma}}{C^{2}} d\theta^{2}
    \label{eq:6}
\end{equation}

\noindent where $\sigma$ and $C$ are real numbers ($C>0$). For some
parameter values one can interpret the Levi-Civita spacetime as the
spacetime of an ``infinite line mass". In fact, after some controversy
in the literature (see, e.g. [\refcite{bonnor}, \refcite{HRS}, \refcite{HSTW}]), the
following interpretations have become somewhat accepted: $\sigma = 0, 1/2$
locally flat; $\sigma =0, \\C=1$ Minkowski spacetime; $\sigma =0, C\not= 1$
cosmic string spacetime; $0 < \sigma < 1/2$ ``infinite line mass" spacetime
(modelled by a scalar curvature singularity at $r=0$); $\sigma = 1/2$ Minkowski 
spacetime in accelerated coordinates (planar source).

\section{Classical and Quantum Singularities}
The analysis in [\refcite{KHW}] uses Weyl's
limit point-limit circle criterion [\refcite{RS}] to determine essential
self-adjointness of the spatial portion of the Klein-Gordon
wave operator on a $C_{0}^{\infty}$ domain in $L^{2}$, a Hilbert space of square integrable functions.
The conclusions will now be summarized.

If $\sigma$ is neither zero nor one-half, the
Klein-Gordon operator is not
essentially self-adjoint,  so all $\sigma \neq
0, \sigma \neq 1/2$ Levi-Civita spacetimes are quantum mechanically
singular as well as being classically singular with scalar curvature singularities.

\par If $\sigma = 0$ and $C = 1$, the spacetime is simply Minkowski
space. One of the two solutions of the radial Klein-Gordon equation
can be rejected because it diverges at a regular point ($r=0$) of the
spacetime. The operator is therefore quantum
mechanically nonsingular (a well known fact, repeated here for
completeness).

\par If $\sigma = 0$ and $C \neq 1$, the spacetime is the conical
spacetime corresponding to an idealized cosmic string. The cosmic string spacetimes
are quantum mechanically
singular for azimuthal quantum number $m$ such that $|m| C < 1$ and
nonsingular if  $|m| C \ge 1$. If arbitrary values of $m$ are allowed,
these
spacetimes are quantum mechanically singular in agreement with earlier
results [\refcite{HK}]. These spacetimes are also
classically singular with a quasiregular (``disclination'')
singularity at $r=0$.

\par If $\sigma = 1/2$ the classical spacetime is flat and without a
classical singularity. This spacetime is also quantum mechanically nonsingular.
The Weyl limit point-limit circle
techniques used in [\refcite{KHW}] emphasize the flatness of the spacetime
and support a description given in [\refcite{bonnor}] of this spacetime as one
given by a cylinder whose  radius has tended to infinity.

\par  For the Levi-Civita spacetimes, all that are classically
singular are also quantum mechanically singular, and all that are
classically nonsingular ($\sigma = 0,\, C = 1$, and $\sigma = 1/2$) are
also quantum mechanically nonsingular.  The classically and
quantum-mechanically nonsingular spacetimes correspond to isolated
values of $\sigma$, so that (for example) even though the spacetime
$\sigma = 0,\, C = 1$ is nonsingular, the spacetimes with  $\sigma
\to 0,\, C = 1$ are singular.  The only discrepency between
classical and quantum singularities are for the $\sigma =0,\, C \neq 1$
modes with $|m| C \ge 1$, which are quantum mechanically nonsingular
in a classically singular spacetime. The physical reason is that the
wavefunction for large values of $m$ in a flat space with a quasiregular
singularity at $r=0$ is unable to detect the presence of the
singularity because of a repulsive centrifugal potential.  

\section{Conclusions}
The limit point-limit circle
criterion that was used in [\refcite{KHW}] provides physical insight into
when quantum singularities are prevented from occurring by potential
barriers as well as the true meaning of the $\sigma = 1/2$ case.

\section*{Acknowledgements}
We thank David Clarke, Jack Harrison and Cassidi Reese for useful
conversations.
One of us (DAK) was partially funded by NSF grants PHY-9988607 and PHY-0241384 to
the U.S. Naval Academy. She also thanks Queen Mary, University of London,
where some of this work was carried out.

\end{document}